\documentclass{article}

\usepackage{arxiv}

\usepackage[utf8]{inputenc} 
\usepackage[T1]{fontenc}    
\usepackage{hyperref}       
\usepackage{url}            
\usepackage{booktabs}       
\usepackage{amsfonts}       
\usepackage{nicefrac}       
\usepackage{microtype}      

\usepackage{graphicx}
\usepackage{amsmath,amssymb,amsfonts, epsf}
\usepackage{algorithmic}
\usepackage[ruled,vlined]{algorithm2e} 
\usepackage{breqn}
\graphicspath{{images/}}
\usepackage{nomencl}
\usepackage{hyperref}
\hypersetup{
    colorlinks=true,
    linkcolor=blue,
    filecolor=magenta,      
    urlcolor=cyan,
}
\usepackage{xcolor}
\usepackage{caption}
\title{Fusing Physics-based and Deep Learning Models for Prognostics}

\author{Manuel Arias Chao \\
	ETH Zurich\\
    \texttt{manuel.arias@ethz.ch} \\
   \And
    Chetan Kulkarni\\
	KBR, Inc., NASA Ames Research Center\\
	\texttt{chetan.s.kulkarni@nasa.gov}\\
   \And
   Kai Goebel\\
   Lule{\aa} University of Technology\\
   \texttt{kai.goebel@ltu.se}\\
   \And	
   Olga Fink \\
   ETH Zurich\\
   \texttt{ofink@ethz.ch}\\
}

\begin{document}
\maketitle

\begin{abstract}

Physics-based and data-driven models for remaining useful lifetime (RUL) prediction typically suffer from two major challenges that limit their applicability to complex real-world domains: (1) incompleteness of physics-based models and (2) limited representativeness of the training dataset for data-driven models. Combining the advantages of these two directions while overcoming some of their limitations, we propose a novel hybrid framework for fusing the information from physics-based performance models with deep learning algorithms for prognostics of complex safety-critical systems under real-world scenarios. In the proposed framework, we use physics-based performance models to infer unobservable model parameters related to a system's components health solving a calibration problem. These parameters are subsequently combined with sensor readings and used as input to a deep neural network to generate a data-driven prognostics model with physics-augmented features. The performance of the hybrid framework is evaluated on an extensive case study comprising run-to-failure degradation trajectories from a fleet of nine turbofan engines under real flight conditions. The experimental results show that the hybrid framework outperforms purely data-driven approaches by extending the prediction horizon by nearly 127\%. Furthermore, it requires less training data and is less sensitive to the limited representativeness of the dataset compared to purely data-driven approaches.
\end{abstract}


\section{Introduction} \label{sec:Introdution}

The prediction of the failure time of complex systems has traditionally been addressed on the basis of models that capture the physics of failure. While extensive research on physics-based models has been performed \cite{Bolander2009, Daigle2011, Daigle2013}, physical degradation processes are only well-understood for critical or relatively simple components. As a result, implementations of physics-based models in practical applications have been limited. 

The increased availability of system condition monitoring data has driven the increased use of data-driven approaches for prognostics and health management (PHM) of complex engineered systems. The underlying assumption of data-driven approaches is that the relevant information of the evolution of the system health and the failure time can be learned from past data \cite{Schwabacher2007}. In particular, deep learning has recently gained attention due to its ability to learn fault patterns more easily directly from raw sensor data \cite{Khan2018}. A variety of supervised and semi-supervised deep learning models have shown promising performance in estimating the RUL from sensor data \cite{Malhotra, Yoon2017, Zhao2017,  Li, Pasa2019, DeOliveiradaCosta2019, LISTOUELLEFSEN2019240, SHI2021107257} based on some prognostics benchmark datasets \cite{Agogino2007, Saxena2008}. Most research studies on deep learning applications in PHM require a representative dataset of end-to-end (i.e., run-to-failure) degradation trajectories to obtain accurate prognostics models. These trajectories need to be comprised of a set of time series sensor readings along with the corresponding time-to-failure labels. However, the collection of such a representative dataset for systems subjected to periodic maintenance interventions can take a long time because (a) failures may be rare and (b) the system can operate in different environments and follow different mission profiles resulting in a large range of possible deterioration trajectories. In real application scenarios, the available datasets generally contain only a small number of units and failure modes and are, therefore, not fully representative of all potential future degraded system conditions. Moreover, for complex engineered systems with a substantial variability of operating conditions and continuously increasing degradation, data-driven approaches struggle to distinguish between the impact of changes in operating conditions and the impact of degradation on the sensor readings. Consequently, data-driven methods have difficulties to relate the condition monitoring data to the asset failure time, limiting their practical applications.

Although data-driven and physics-based approaches have limitations when applied as stand-alone approaches, it is hypothesized here that the combined use of data-driven and physics-based approaches can potentially lead to performance gains by leveraging the advantages of each method. In particular, while physics-based approaches are generally limited by the inability to properly tune parameters of models with high complexity or model incompleteness, they do not require large amounts of data, retain the interpretability of a model, and one can generate synthetic data. In contrast, data-driven approaches are limited by the representativeness of the training datasets but are simple to implement, and one can use data-driven models to discover complex patterns from large volumes of data. It follows that data-driven solutions can be advantageous to enhance or replace inaccurate parts of the physics-based models. In addition, physics-based model information can help to reduce the required amount of training data (i.e., overcoming the lack of time-to-failure trajectories) by generating synthetic data or providing model parameters that are very informative for data-driven models. In general, the physics-based system models can be used as 'teachers' to guide the discovery of meaningful machine learning models.

Various approaches have been proposed to combine physics-based and data-driven approaches. Depending on what type of information is processed and how the pieces of information are combined, different types of hybrid architectures can be created \cite{Rai2020, willard2020integrating}. Some examples of hybrid models for prognostics are \cite{Liao2014, Pillai2016, Zhang2017, Nascimento2019, Dourado2019, Yucesan2019}. In particular, hybrid systems combining thermodynamic performance models and data-driven aging models have shown promising results on individual simpler systems such as lithium batteries \cite{Zhang2017}. Recently, several approaches of physics-guided machine learning have been proposed, where physical principles are used to inform the search for a physically meaningful and accurate machine learning model. The architecture proposed in \cite{Jia2018}, for example, enhances the input space to a data-driven system model with outputs from a physics-based system model. As a result, the authors showed how the dynamical behavior of the system could be approximated more accurately. In another variation of the physics-guided machine learning idea, a recurrent neural network (RNN) cell was modified to incorporate the information from the system model as an internal state of the RNN. A related idea was applied to a variety of prognostics problems, such as in \cite{Nascimento2019, Dourado2019, Yucesan2019}. The underlying hypothesis of this method is that the output of the physics-based model is informative of the degradation process and, consequently, of the failure time. 

\begin{figure}[ht]
\centering
\includegraphics[width=9.0cm]{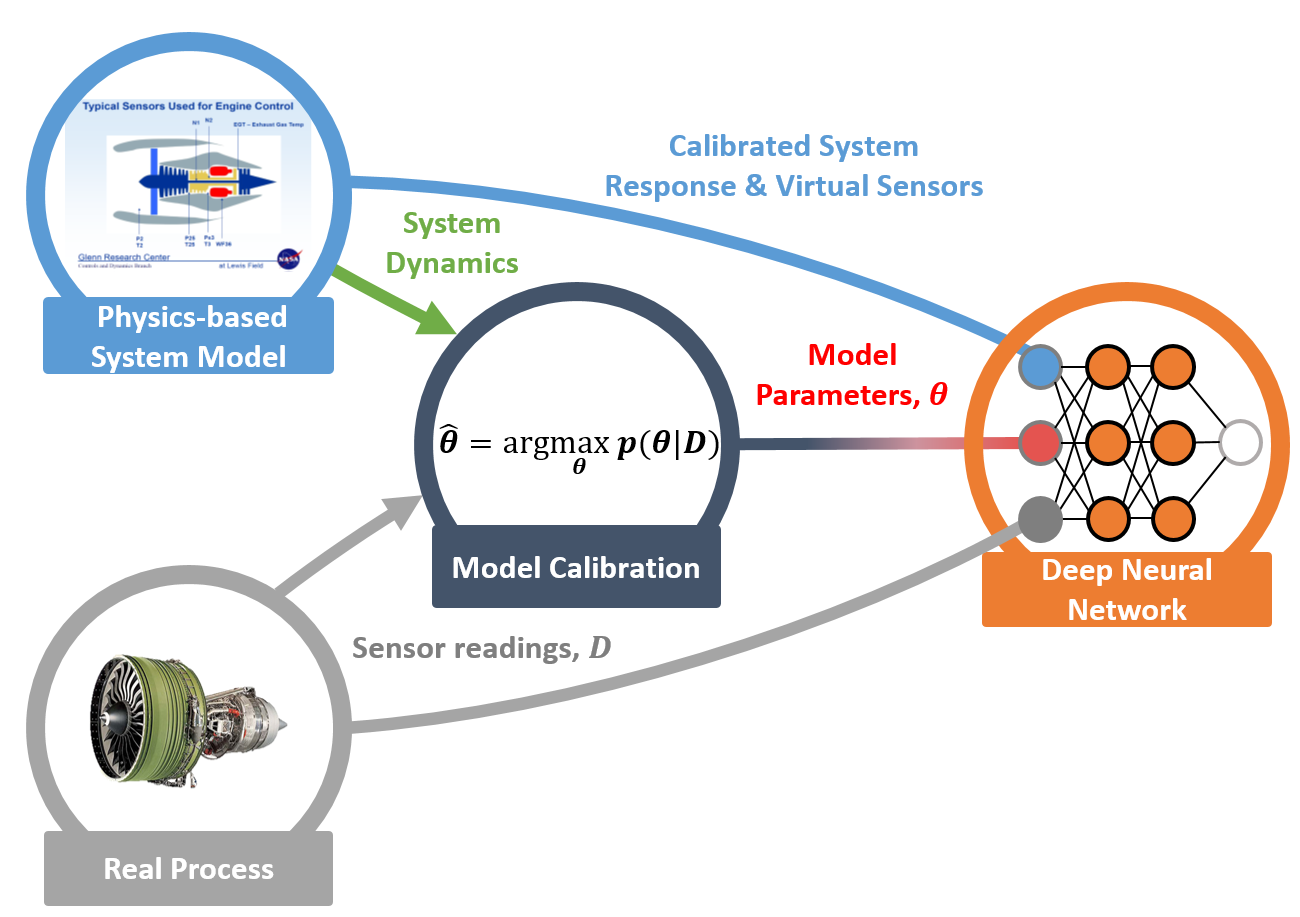}
\caption{Proposed hybrid prognostics framework fusing physics-based and deep learning models. Given the system dynamics and sensor readings, we perform the calibration of the system model to estimate unobservable model parameters $\theta$ that encode the health condition of the system component. These parameters are subsequently combined with sensor readings from the condition monitoring (CM) data and calibrated system model responses containing unobserved process properties (i.e., virtual sensors) and used as input to a deep neural network to generate the deep learning-based prognostics model.}
\label{fig:trailer}
\end{figure}

In contrast to the hybrid architectures cited above, the framework presented in this paper leverages inferred \textbf{unobserved virtual sensors} and \textbf{unobservable parameters} of physics-based system models closely related to the system health to enhance the input space to deep learning-based prognostics models. For the physics-based system models, we focus on thermodynamic performance models (0D models) that are generally available for design, control, or performance evaluation of complex systems. Using as inputs the modeled system dynamics and the sensor readings from the condition monitoring data, we infer model parameters (e.g., efficiencies and flows modifiers) and unobserved process properties (e.g., temperatures and pressures) that are informative of the health condition and its evolution in time by solving a calibration problem. The model parameters and process properties (i.e., virtual sensors) are subsequently combined with sensor readings and used as input to a deep neural network to generate a data-driven prognostics model. An overview of the proposed framework is shown in Figure \ref{fig:trailer}. A calibration-based hybrid framework was applied in \cite{AriasChao2019} to diagnostics problems. In this paper, we notably extend the framework to address the problem of remaining useful life estimation.

The performance of the proposed hybrid framework is evaluated on a synthetic dataset comprising a small fleet of nine turbofan engines with run-to-failure degradation trajectories exhibiting a high variability in operating conditions. A scenario of incomplete representation of the test degradation conditions in the training dataset is considered. The dataset was generated with a modified version of the Commercial Modular Aero-Propulsion System Simulation (C-MAPSS) dynamical model \cite{Frederick2007}. Real flight conditions as recorded onboard commercial jets were taken as input to the C-MAPSS model \cite{DASHlink}. The performance of the approach is compared to an alternative data-driven approach where only sensor data are used as input to two types of deep neural networks (a multilayer perceptron feed-forward neural networks and Convolutional Neural Networks (CNN)). The proposed hybrid method outperforms the equivalent data-driven approaches and provides superior results in RUL estimation under highly varying operating conditions and incomplete representation of the training dataset. Furthermore, the hybrid framework demonstrates to require less training compared to the purely data-driven algorithms. 

The remainder of the paper is organized as follows. In section \ref{sec:background} the background of the solution strategy is provided, and the prognostics problem is formally introduced. In Section \ref{sec:Methods} the proposed framework is described.  Section \ref{sec:Case_study}, introduces the case study. In section \ref{sec:Results} the results are presented. Finally, a summary of the work and outlook are given in Section \ref{sec:Conclusion}.

\section{Background} \label{sec:background}

This section briefly introduces the basic concepts and notations related to system performance models and calibration of physics-based models as they are the building blocks of the proposed framework.

\subsection{System Performance Models}  \label{sec:background_system_models}
The design of complex engineered systems involves modeling the physical principles governing the system performance and the thorough validation of those models with field data. Hence, thermodynamic performance models with different fidelity levels are typically available for control and performance evaluation of complex systems \cite{Roth2005, AriasChao2015}. These system models have, typically, a moderate computational load and are yet able to predict measured process variables (e.g., temperatures, pressures, or rotational speeds) as well as global unmeasured system and sub-system performance (e.g., efficiencies and power). Since, in the general case, there is no description given by an explicit formula, the thermodynamic performance models are represented mathematically as coupled systems of nonlinear equations. The inputs of the performance models are divided into scenario-descriptor operating conditions $w$, and unobservable model parameters $\theta$. The unobservable model parameters $\theta$ correspond to tuning parameters generally related to the health condition of the sub-components of the system. The outputs of the model are estimates of the measured physical properties $\hat{x}_s$ and unobserved properties $\hat{x}_v$ that are not part of the condition monitoring signals (i.e., \textit{virtual sensors}). Hence, the nonlinear performance model is denoted as:
\begin{equation}
  [\hat{x}_s,\hat{x}_v] = F(w,\theta)
\end{equation}

\subsection{Calibration of Physics-Based Models}  \label{sec:background_calibration_methods}
Inference of system model parameters from observations $x_s$ is often referred to as calibration \cite{Kennedy2001}. System model calibration is an inverse problem aiming at obtaining the values of the model parameters $\theta$ that make the system response follow the observations, i.e., $\hat{x}_s \sim x_s$. Therefore, the problem of calibration of physics-based models corresponds to the problem of modeling a physical process as approximated by a physics-based model. Since both the observations and model parameters are uncertain, model calibration is a stochastic problem. Ideally, the calibration process aims at obtaining the posterior distribution of the calibration factors given the data $p(\theta|w, x_s)$. However, computing the whole distribution is generally computationally expensive and, therefore, in most cases, point value estimations of the parameters are inferred. A typical compromise is to compute the \emph{maximum a posteriori estimation} (MAP), described by
\begin{equation} \label{eq:map}
 \hat{\theta}_{\text{MAP}} = \arg\max_\theta p(\theta|w,x_s)
\end{equation}
Several methods have been proposed to address the problem of dynamical model calibration when the physics-based model structure is well-founded on known physical principles (e.g., aircraft thermodynamic engine models). The majority of the available methods are probabilistic, or estimation approaches developed in the fields of statistics \cite{Sacks1989} and optimal control \cite{Crassidis2011}. Some examples of popular estimation methods include iterative reweighted least-squares schemes \cite{AriasChao2015}, unscented Kalman filters (UKF) \cite{Julier1997, Turner2010, Borguet2012}, particle filters \cite{Kantas2015} or Bayesian inference methods using Markov chain Monte Carlo \cite{Rutter2009, AriasChao2015}. Recently, deep learning methods based on reinforcement learning and direct mappings with supervised learning have also been proposed in \cite{tian2020realtime}.

\subsection{Problem Formulation} \label{sec:background_formulation}
Given are multivariate time-series of condition monitoring sensors readings $x_{s_i} = [x_{s_i}^{(1)}, \dots, x_{s_i}^{(m_i)}]^T$ and their corresponding RUL i.e., $y_i=[y_i^{1},\dots, y_i^{m_i}]^T$ from a fleet of $N$ units ($i=1, \dots, N$). Each observation $x_{s_i}^{(t)} \in R^{p}$ is a vector of $p$ raw measurements taken at operating conditions $w_i^{(t)} \in R^{s}$. The length of the sensory signal for the \textit{i-th} unit is given by $m_{i}$; which can, in general, differ from unit to unit. The total combined length of the available data set is $m={\sum_{i=1}^{N}m_i}$. More compactly, we denote the available dataset as $\mathcal{D} =\{w_i, x_{s_i}, y_i\}_{i=1}^{N}$. In addition to the CM data and the RUL label, we have access to a performance system model $F(w,\theta_{\text{ref}})$ which provides the expected dynamical response of a non-degraded reference unit (i.e., $\theta = \theta_{\text{ref}}$). Starting from an unknown initial health condition, the CM data of each unit records the degradation process of the system's components. The system's components experience \emph{normal} (linear) degradation until point in time $t_{s_i}$ where an \emph{abnormal} condition arises leading to an eventual failure at $t_{\text{EOL}_i}$ (end-of-life).

Given this set-up, the task is to obtain a predictive model $\mathcal{G}$ that provides a reliable RUL estimate ($\mathbf{\hat{y}}$) on a test dataset of $M$ units $\mathcal{D}_{T*}=\{x_{s_j*}\}_{j=1}^{M}$; where $x_{s_j*} = [x_{s_j*}^{1},\dots, x_{s_j*}^{k_j}]$ are multivariate time-series of sensors readings. The total combined length of the test data set is $m_*={\sum_{j=1}^{M}k_j}$.

\section{Proposed Framework: Deep Learning-Based Prognostics with Physics-Inferred Inputs} \label{sec:Methods}

In this work, we propose to combine a calibrated physics-based performance model with deep learning architectures to obtain accurate hybrid prognostics models.

The calibration of physics-based system models provides estimates of the model parameters ($\hat{\theta}$) that encode and explain the deteriorated behavior of the sub-components. The resulting calibrated model, i.e., $F(w,\hat{\theta})$, provides high confidence estimates of unobserved process variables $\hat{x}_v$ that may be sensitive to fault signatures. As a result, model calibration increases the amount of information available for developing a data-driven prognostics model. Although the physics-inferred information can be combined with the CM data in multiples ways, in this work, we propose to enhance the input feature space of a data-driven prognostics model with the process variables $[\hat{x}_s, \hat{x}_v, \hat{\theta}]$ to develop a hybrid prognostics model. In particular, to benefit from the learning ability of recent advances in deep learning, we propose to combine the physics-based performance models with deep learning architectures.     

\begin{figure}[ht]
\centering
\includegraphics[width=15.0cm]{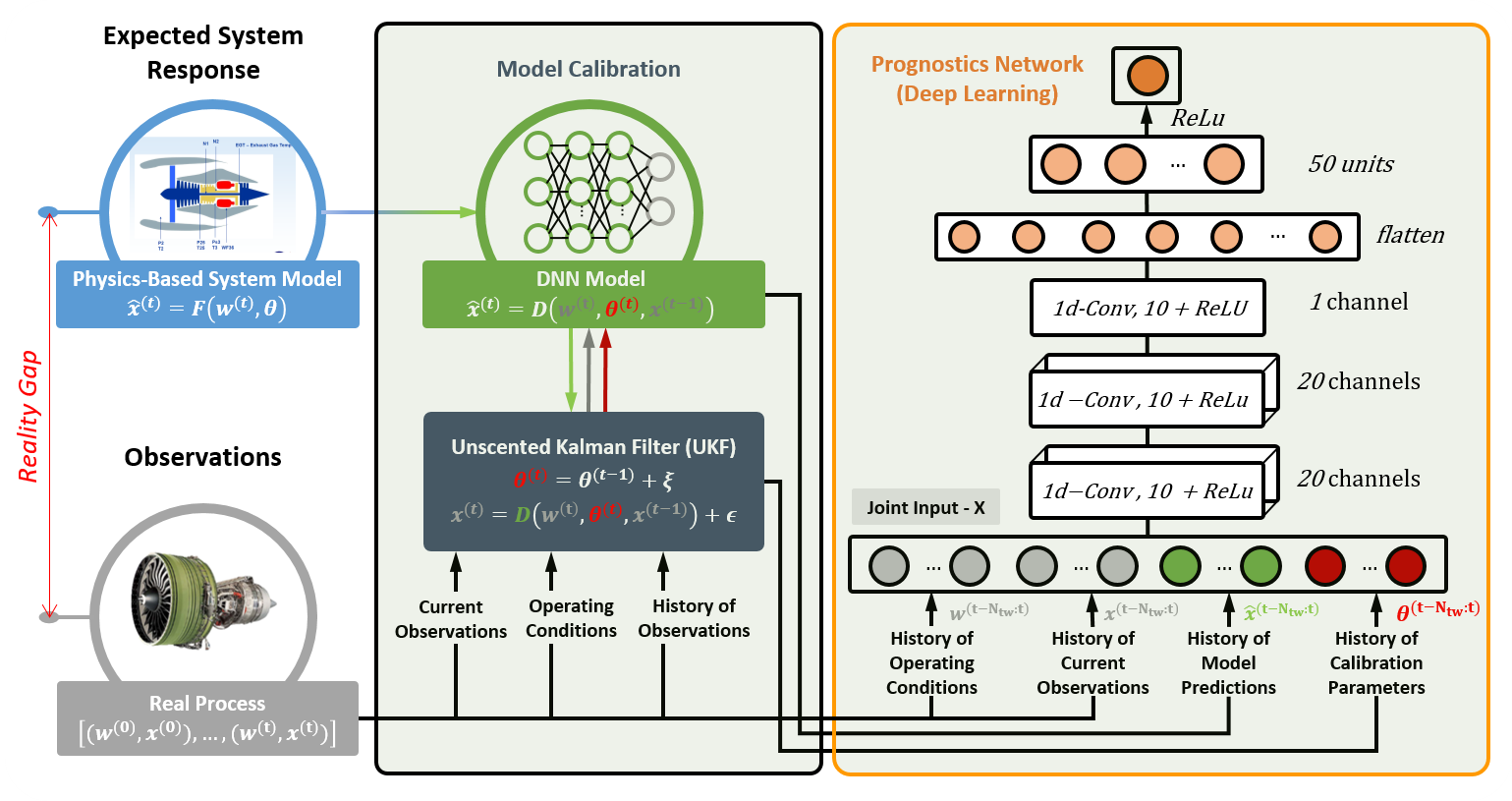}
\caption{Overview of the hybrid prognostics framework fusing physics-based and deep learning models. The deep learning prognostics model receives as input the scenario-descriptor operating conditions ($w$), estimates of the condition monitoring signals ($\hat{x}_s$), and the virtual sensors ($\hat{x}_v$) and unobservable model parameters ($\hat{\theta}$). The calibration of the system model (i.e., the estimation of the unobservable model parameters $\theta$) is carried out with a state-space formulation using a UKF. We use a discrete-time counterpart of the physics-based model $F$ to speed up the calibration process. The dynamical system $D$ is modeled by a deep neural network. A convolutional neural network (CNN) architecture is depicted as deep learning-based prognostics model (see Section \ref{sec:networks} for further details)}
\label{fig:hybrid_cal}
\end{figure}

Figure \ref{fig:hybrid_cal} shows a block diagram of the proposed calibration-based hybrid prognostics approach. The deep learning prognostics model receives scenario-descriptor operating conditions $w$ and model variables $[\hat{x}_s, \hat{x}_v, \hat{\theta}]$ as input. Contrarily to a standard data-driven approach aiming to learn a mapping function from the CM signals to the RUL target $\mathbf{y}$ (i.e., $[\mathbf{w}, \mathbf{x}_s] \longmapsto \mathbf{y}$), we first obtain a more informative representation $\mathbf{x}$ with additional health related features inferred by calibration of a physics-based system model ($[\mathbf{x_s}, F(\mathbf{w},\boldsymbol{\cdot})]\longmapsto \hat{\theta}$). In a second step, we find an optimal mapping $\mathbf{\mathcal{G}}: \mathbf{x}\longmapsto \mathbf{y}$ from the enhanced input space $\mathbf{x}=[w, x_{s}, \hat{x}_{s}, \hat{x}_{v}, \hat{\theta}]$ to perform RUL estimation. The hybrid framework is very flexible and can be combined with any type of calibration method and deep learning architectures.

The proposed hybrid methodology provides the following advantages compared to pure data-driven approaches: 

\begin{itemize}
 \item Ability to accurately predict the remaining useful lifetime also when available datasets to train data-driven approaches are sparse.
  \item Interpretability of the model and the corresponding outputs in terms of physically meaningful input features.
  \item Robustness to sensor faults that can be distinguished against faulty conditions
  \item Reduction of the required size of training dataset for similar or better performance
  \item Ability to generate additional operating conditions to compensate for the lacking CM data 
\end{itemize}

\subsection{Calibration of the System Performance Model} \label{sec:methods-calibration}
In this work, instead of focusing on one particular model calibration method, we are aiming at demonstrating the benefits of combining physics-inferred model parameters representing sub-model health with deep learning models to generate accurate prognostics models. Furthermore, the goal is to evaluate the impact of different levels of calibration accuracy on the performance of the proposed prognostics framework. Since the calibration of the performance model itself is not in the focus of this research, we apply a state-of-the-art approach for calibration: an Unscented Kalman Filter \cite{Julier1997} to infer the values of the model correcting parameters $\theta$. The rationale for this choice is that our models of interest are nonlinear and that UKF provides a good compromise between computational cost and performance. In fact, UKF is widely applied for aircraft engine health evaluation \cite{Crassidis2011, Borguet2012, AriasChao2019}. However, we would like to stress that the proposed framework is flexible to the chosen model calibration approach. 

Model parameter estimation with UKF involves a traditional state-space formulation. In this solution strategy, the state vector comprises the health parameters; which are modelled as a random walk. The measurement equation depends on the states and the input signals at the present time step $t$. The measurement equation is readily available from the system model $F$. Hence, we consider a nonlinear discrete time system of the form:
\begin{align} \label{eq:ukf}
      \theta^{(t)} &= \theta^{(t-1)} + \xi^{(t)}  \\
       x_s^{(t)} &= F(w^{(t)}, \theta^{(t)}) + \epsilon^{(t)}
\end{align}
where $\xi \sim N(0,Q)$ is a Gaussian noise with covariance $Q$ and $\epsilon \sim N(0,R)$ is a Gaussian noise with covariance $R$. A more detailed explanation of this problem formulation applied to the monitoring of gas turbine engines can be found in \cite{Borguet2012}.  

In order to speed up the learning process of the UKF algorithm, a discrete time counterpart of the physics-based model $F$ is used. The resulting dynamical system $D$ or simulator is modelled by a deep neural network that approximates the dynamic transition equation describing how the expected system response changes given the current observations $x^{t}$, the control variables $w^{t+1}$, and model parameters $\theta^{t+1}$, resulting in:
\begin{align} \label{eq:discrete_system}
 \hat{x}^{(t)}=D(w^{(t)}, \hat{x}^{(t-1)}, \theta^{(t)}) 
\end{align}

\subsection{Deep Learning-Based Prognostics with Physics Inferred Inputs} \label{sec:calibration-methods}
The inferred model parameters $\hat{\theta}$ are informative about the health state of the system but do not directly relate to the end-of-life time. Together with the scenario-descriptor operating conditions $w$, they comprise the 'expected' independent factors of variation of sensor readings ($x_s$). Consequently, the inferred model parameters $\hat{\theta}$ are ideal parameters to disentangle the contribution of system degradation and operation condition change $w$ from the observed system responses. Following this reasoning, the model correcting parameters are the perfect complement to the raw sensor readings for the generation of data-driven prognostics models.

Since deep learning models have shown an excellent ability to reveal hidden complex functional mapping between inputs and target labels, we choose a deep neural network to discover a mapping $\mathcal{G}$ that relates the enhanced input  $x=[w, \hat{x}_{s}, \hat{x}_{v}, \hat{\theta}]$ to a target label $\mathbf{y}$ given a training set $S_{T} \subsetneq \mathcal{D}$. Again, multiple learning strategies are possible for this task (supervised or semi-supervised learning). In this research, we chose the standard supervised learning strategy (SL), i.e., a $\mathbf{direct}$ mapping from input $\mathbf{x}$ to a target label $\mathbf{y}$. The main reasons for this choice are the simplicity and suitability to the problem formulation in Section \ref{sec:background_formulation}. It should be pointed out that under a slightly different problem formulation, a semi-supervised learning or a domain adaptation strategy could also be potentially applied. To obtain the mapping function $\mathcal{G}$, a deep multilayer perceptron feed-forward neural network (\text{FNN}) and a convolutional neural network (CNN) are evaluated within the proposed framework. As already mentioned earlier, different types of architectures, including also recurrent structures, could be used within the proposed framework. Section \ref{sec:networks} provides further details about the architecture. The entire procedure proposed in this paper is summarized in Algorithm \ref{al:cbhdlp}. 

\begin{algorithm}[ht]
\caption{Calibration-based Hybrid Deep Learning Prognostics} \label{al:cbhdlp}
\SetAlgoLined
\DontPrintSemicolon

\textbf{Input}: $\{w^{(i)}, x^{(i)}_s\}_{i=1}^m \in \mathcal{D}$ \& $F(w^{(t)},\theta_{\text{ref}})$    \\
\For{$i = 1:m$}{
$\hat{\theta}^{(i)} \gets \arg\max_\theta p(\theta^{(i)}|(w^{(i)}, x^{(i)}_s)$         \\
}

\textbf{Input}: $\{w^{(i)}, x^{(i)}_s, \hat{x}^{(i)}_s, \hat{x}^{(i)}_v, \hat{\theta}^{(i)}, y^{(i)}\}_{i=1}^{m_{S_T}} \in  S_T$ \\
$X = \{w^{(i)}, x^{(i)}_s, \hat{x}^{(i)}_v, \hat{\theta}^{(i)}\}_{i=1}^{m_{S_T}}$ Enhanced Input -$X$
\While{$i \leq E_s$}{
$\mathcal{H} \gets \text{Update parameters in Prognostics Network using SGD}$                                 \\
}

\textbf{Input}: $\{w^{(j)}_*, x^{(j)}_{s*}\}_{j=1}^{m_*} \in \mathcal{D}_{T*}$ \& $F(w^{(t)},\theta_{\text{ref}})$ \\
\For{$j = 1:m_{*}$}{
$\hat{\theta}^{(j)}_* \gets \arg\max_\theta p(\theta^{(j)}|(w^{(j)}_*, x^{(j)}_{s*}))$        \\
}
\textbf{Input}: $\{w^{(j)}_*, x^{(j)}_{s*}, \hat{x}^{(j)}_{v*}, \hat{\theta}^{(j)}_*\}_{j=1}^{m_*} \in D_T$\\
$X_* = \{w^{(j)}_*, x^{(j)}_{s*}, \hat{x}^{(j)}_{v*}, \hat{\theta}^{(j)}_*\}_{j=1}^{m_*}$ Enhanced Input -$X_*$  \\
\For{$j = 1:m_{*}$}{
$\hat{y}^{(j)} \sim \mathbf{\mathcal{G}}(x^{(j)}_*;\mathcal{H})$                         \\
}
\end{algorithm}

\section{Case Study} \label{sec:Case_study}

\subsection{A Small Fleet of Turbofan Engines}
\begin{figure}[h]
\centering
\includegraphics[width=7.5cm]{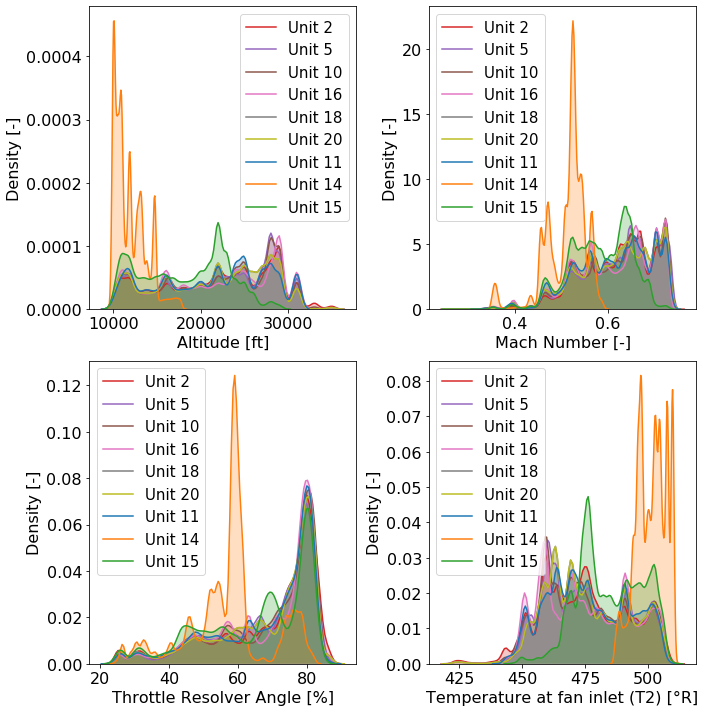}
\includegraphics[width=7.5cm]{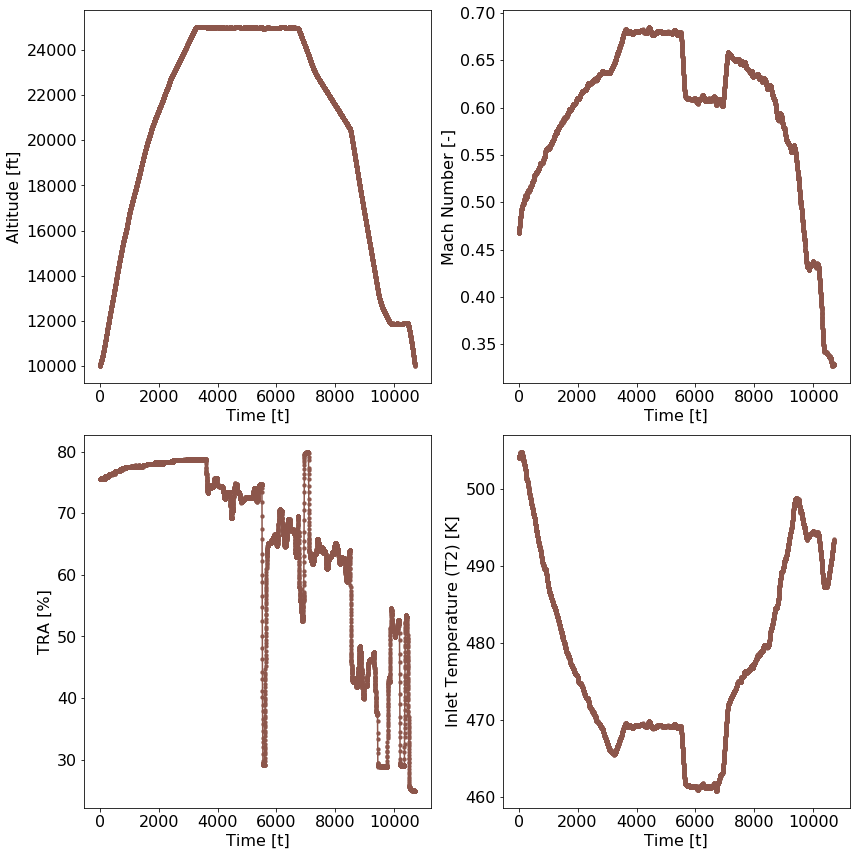}
\caption{\textbf{Left.} Kernel density estimations of the simulated flight envelopes (i.e., climb, cruise and descend flight conditions) given by recordings of altitude, flight Mach number, throttle-resolver angle (TRA) and total temperature at the fan inlet (T2) for complete run-to-failure trajectories of six training units ($u=$ 2, 5, 10, 16, 18 \& 20) and three test units ($u=$ 11, 14 \& 15). \textbf{Right.} An example of a typical single flight cycle given by traces of the scenario-descriptor variables for Unit 10. Climb, cruise and descend flight conditions (with $\text{alt}>10000$ ft) corresponding to different flight routes operated by the aircraft are covered.}
\label{fig:operation_hist}
\end{figure}

The proposed methodology is demonstrated and evaluated on a synthetic dataset with run-to-failure degradation trajectories of a small fleet comprising nine turbofan engines with unknown and different initial health conditions. The dataset was generated with the Commercial Modular Aero-Propulsion System Simulation (C-MAPSS) dynamical model \cite{Frederick2007}. Real flight conditions as recorded onboard of commercial jets were taken as input to the C-MAPSS model \cite{DASHlink}. Figure \ref{fig:operation_hist} (left) shows the kernel density estimations of the simulated flight envelopes given by the scenario-descriptor variables $W$: altitude ($\text{alt}$), flight Mach number ($\text{XM}$), throttle-resolver angle ($\text{TRA}$) and total temperature at the fan inlet ($\text{T2}$) for $N=6$ training units ($u=$ 2, 5, 10, 16, 18 \& 20) and $M=3$ test units ($u=$ 11, 14 \& 15). It is worth noticing that test units 14 and 15 have an operation distribution significantly different from training units. Concretely, test units 14 and 15 operate shorter and lower altitude flights compared to other units. The training dataset contains, thereby, flight profiles that are not fully representative for the test conditions of these two units. This is a more difficult learning and generalization task. We have chosen this example due to its relevance for practical applications where observed operating conditions of new units may not correspond to the past operating conditions of other units in the fleet. Purely data-driven approaches generally require domain-adaptation approaches for this type of setups \cite{wang2020missing}. 

\begin{figure}[ht]
\centering
\includegraphics[width=9.5cm]{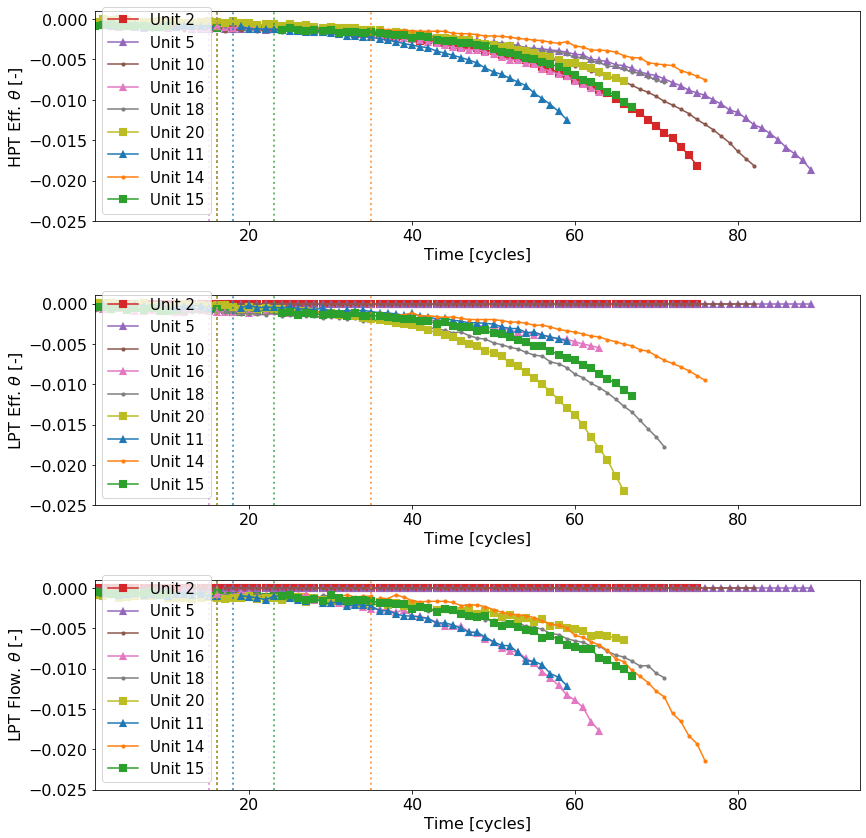}
\caption{Traces of the degradation imposed on the high pressure turbine efficiency (HPT\_Eff\_mod), low pressure turbine efficiency (LPT\_Eff\_mod) and low pressure turbine flow (LPT\_flow\_mod) for each unit of the fleet. The onset of the \text{abnormal degradation} (i.e. $t_{s}$) of each unit is indicated by dashed vertical lines.}
\label{fig:degradation_units}
\end{figure}

Two distinctive failure modes are present in the available dataset ($\mathcal{D}$). Units 2, 5, and 10 have failure modes of an \emph{abnormal} high-pressure turbine (HPT) efficiency degradation. Units 16, 18, and 20 are subject to a more complex failure mode that affects the low-pressure turbine (LPT) efficiency and flow in combination with the high-pressure turbine (HPT) efficiency degradation. Test units are subjected to the same complex failure mode. Figure \ref{fig:degradation_units} shows degradation profiles induced in the nine units of the fleet. The initial deterioration state of each unit is different and corresponds to an engine-to-engine variability equivalent to 10\% of the health index. The degradation of the affected system components follows a stochastic process with a linear \emph{normal degradation} followed by a steeper \emph{abnormal degradation}. The degradation rate of each component varies within the fleet. The transition from \emph{normal} to \emph{abnormal} degradation is smooth and occurs at different cycle times for each unit. The transition time ($t_{s}$) is dependent on the operating conditions, i.e., flight and degradation profile. It should be noted that although the degradation profiles of individual components show nearly overlapping trajectories, the combined profile, i.e., the profile in the three dimensions is clearly different. More details about the generation process can be found in \cite{AriasChao2020}.

An overview of the transition times $t_{s}$, the end-of-life times $t_{EOL}$ and the number of samples from each unit of the fleet $m_i$ is provided in Table \ref{tb:datasets}. The sampling rate of the data is 0.1Hz resulting in a total size of the dataset of 0.53M samples for model development and 0.12M samples for testing. It is worth noticing that while test unit 14 is a short flight engine with the lowest amount of flight time (0.16M seconds) it has the largest number of flight cycles.    

%
\begin{table}[ht]
  \begin{center}
    \begin{tabular}{|c||c|c|c|c|}
      \hline
      \multicolumn{5}{|c|}{Development Dataset - $\mathcal{D}$}                 \\ \hline \hline
      Unit ($u$)     & $m_i$     & $t_{s}$          & $t_{EOL}$          & Failure Mode \\ \hline   
       2       &   0.85M   &   17             &   75               &  HPT         \\ \hline 
       5       &   1.03M   &   17             &   89               &  HPT         \\ \hline 
       10      &   0.95M   &   17             &   82               &  HPT         \\ \hline
       16      &   0.77M   &   16             &   63               &  HPT+LPT     \\ \hline 
       18      &   0.89M   &   17             &   71               &  HPT+LPT     \\ \hline 
       20      &   0.77M   &   17             &   66               &  HPT+LPT     \\ \hline \hline
      \multicolumn{5}{|c|}{Test Dataset - $\mathcal{D}_{T*}$}                    \\ \hline \hline
      Unit ($u$)    & $m_j$   & $t_{s}$          & $t_{EOL}$          & Failure Mode \\ \hline   
       11      &   0.66M   &   19             &   59               &  HPT+LPT     \\ \hline 
       14      &   0.16M   &   36             &   76               &  HPT+LPT     \\ \hline 
       15      &   0.43M   &   24             &   67               &  HPT+LPT     \\ \hline 
    \end{tabular}
  \end{center}
\caption[Table caption text]{Size ($m_u$), the transition cycle time ($t_{s}$) and end-of-life time ($t_{EOL}$ ) of each unit within the development ($\mathcal{D}$) and test datasets ($\mathcal{D}_{T*}$).}
\label{tb:datasets}
\end{table}

In addition to the condition monitoring (CM) data, we have access to a system model $F(w,\theta_{\text{ref}})$ (i.e., C-MAPSS dynamical model) which provides the expected dynamical response of a non-degraded reference unit ($\theta = \theta_{\text{ref}}$). It is worth noticing that this reference system response deviates from the responses of each of the units due to the different initial health conditions and the degradation trajectories experienced by each unit.

\subsection{Pre-processing}
The dimension of the input space $n$ ($\textbf{x} \in R^{m \times n}$) varies depending on the selected solution strategy. The data-driven models are only based on condition monitoring signals ($[w, x_s]$ and have 20 inputs (i.e. $n = 20$). The proposed hybrid method has 50 inputs (including additionally the model predictions, calibration parameters and the virtual sensors i.e $\textbf{x} = [w, x_s, \hat{x}_s, \hat{x}_v, \hat{\theta}]$). Tables \ref{tb:W_X_m} to \ref{tb:theta} provide a detailed overview of the model variables included in the condition monitoring signals $[w, x_s]$, virtual sensors $x_v$ and model parameters $\theta$. The variable name corresponds to the internal variable name used in CMAPSS model. The descriptions and units are reported as provided in the model documentation \cite{Frederick2007}.


\begin{table}[ht]
\begin{center}
\begin{tabular}{cllc}
\hline
\#    Symbol        &  Description                       & Units          \\ \hline
1    & alt          & Altitude                           & ft             \\
2    & XM           & Flight Mach number                 & -              \\
3    & TRA          & Throttle-resolver angle            & \%             \\
4    & Wf           & Fuel flow                          & pps            \\
5    & Nf           & Physical fan speed                 & rpm            \\
6    & Nc           & Physical core speed                & rpm            \\
7    & T2           & Total temperature at fan inlet     & $^{\circ}$R    \\
8    & T24          & Total temperature at LPC outlet    & $^{\circ}$R    \\
9    & T30          & Total temperature at HPC outlet    & $^{\circ}$R    \\
10   & T40          & Total temp. at burner outlet       & $^{\circ}$R    \\
11   & T48          & Total temperature at HPT outlet    & $^{\circ}$R    \\
12   & T50          & Total temperature at LPT outlet    & $^{\circ}$R    \\
13   & P15          & Total pressure in bypass-duct      & psia           \\
14   & P2           & Total pressure at fan inlet        & psia           \\
15   & P21          & Total pressure at fan outlet       & psia           \\
16   & P24          & Total pressure at LPC outlet       & psia           \\
17   & Ps30         & Static pressure at HPC outlet      & psia           \\
18   & P30          & Total pressure at HPC outlet       & psia           \\
19   & P40          & Total pressure at burner outlet    & psia           \\
20   & P50          & Total pressure at LPT outlet       & psia           \\ \hline
\end{tabular}
\end{center}
\caption[Table tb:CM]{Condition monitoring signals - $[w,x_s]$. The variable symbol corresponds to the internal variable name in CMAPSS. The descriptions and units are reported as in the model documentation \cite{Frederick2007}.}
\label{tb:W_X_m}
\end{table}

\begin{table}[ht]
\begin{center}
\begin{tabular}{cllc}
\hline
\#   & Symbol     &  Description                       & Units        \\\hline
1    & P45        & Total pressure at HPT outlet       & psia         \\
2    & W21        & Fan flow                           & pps          \\
3    & W22        & Flow out of LPC                    & lbm/s        \\
4    & W25        & Flow into HPC                      & lbm/s        \\
5    & W31        & HPT coolant bleed                  & lbm/s        \\
6    & W32        & HPT coolant bleed                  & lbm/s        \\
7    & W48        & Flow out of HPT                    & lbm/s        \\
8    & W50        & Flow out of LPT                    & lbm/s        \\
9    & SmFan      & Fan stall margin                   & --           \\
10   & SmLPC      & LPC stall margin                   & --           \\
11   & SmHPC      & HPC stall margin                   & --           \\
\hline
\end{tabular}
\end{center}
\caption[Table tb:Xv]{Virtual sensors - $[x_v]$. The variable symbol corresponds to the internal variable name in CMAPSS. The descriptions and units are reported as in the model documentation \cite{Frederick2007}.}
\label{tb:X_v}
\end{table}

\begin{table}[hb]
\begin{center}
\begin{tabular}{cllc} \hline
\#  & Symbol         &  Description                       & Units   \\\hline
1   & HPT\_eff\_mod  & HPT efficiency modifier            & -       \\
2   & LPT\_eff\_mod  & LPT efficiency modifier            & -       \\
3   & LPT\_flow\_mod & HPT flow modifier                  & -       \\ \hline
\end{tabular}
\end{center}
\caption[Table tb:failures]{Model correcting parameters - $[\theta]$. The variable symbol corresponds to the internal variable name in CMAPSS. The descriptions and units are reported as in the model documentation \cite{Frederick2007}.}
\label{tb:theta}
\end{table}

The input space $\textbf{x}$ to the models is normalized to a range $[-1, 1]$ by a min/max-normalization given the available dataset ($\mathcal{D}$). A validation set $V_{T} \subsetneq \mathcal{D}$ comprising 10 \% of the available data was chosen for early stopping of the training process and hyperparameter tuning of the deep learning prognostics model. For CNN model requirements, the original dataset was pre-processed with a sliding time window approach of size $N_{tw} = 50$ and stride of 1. The sliding window means that the first input sample to the network takes measurements from timestamps 1-50, the second 2-51, the third 3-52, and so on for each unit of the fleet (i.e., each input has a time length of 500 s).

\subsection{Deep Learning Prognostics Model} \label{sec:networks}
As mentioned before, the main goal of this research study is to evaluate the capability of the proposed hybrid framework to perform prognostics and compare the framework to the performance of purely data-driven approaches.

\begin{figure}[h]
\centering
\includegraphics[width=12cm]{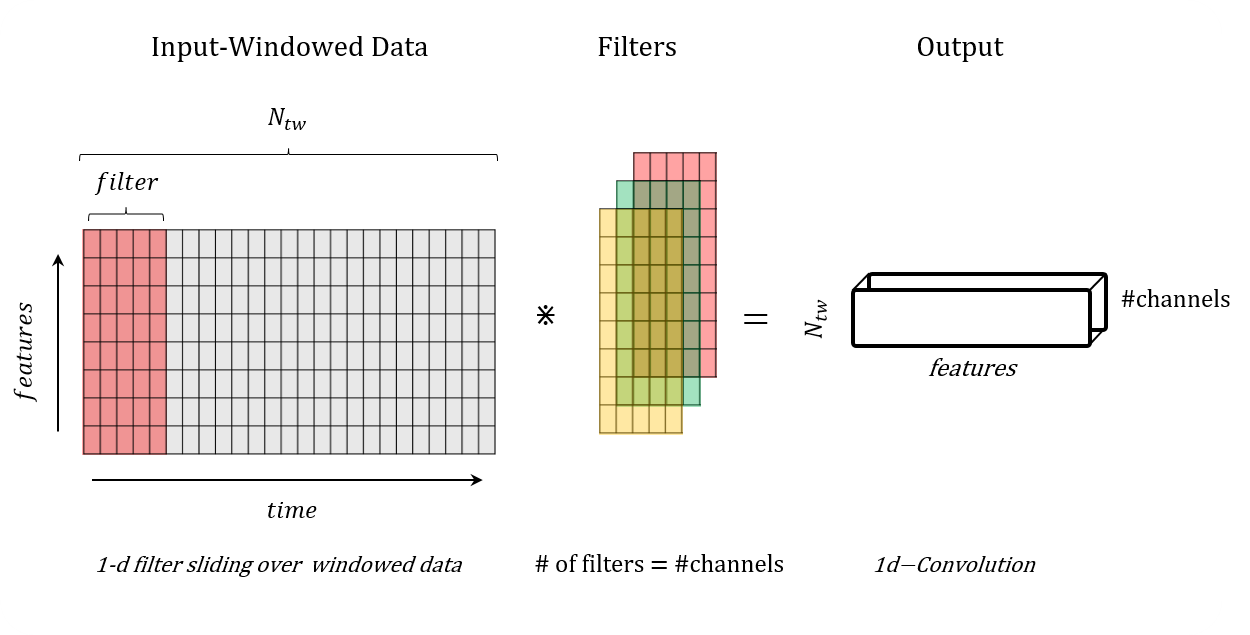}
\caption{1D convolution layer with convolution in the temporal direction }
\label{fig:1-d_conv}
\end{figure}

In prognostics, the degradation mechanisms are often mapped to a health index ($\text{HI}$). This health index can be defined as a normalized margin to multiple health-related thresholds evaluated at specified reference conditions. The end-of-life time of a mechanical system corresponds to the point in time where $\text{HI}=0$. Under this definition, the health index is time-independent and consequently, a mapping from the system state, represented by CM data and $\hat{\theta}$, to $\text{HI}$ exists. While the evolution of the system state is a time-dependent process, the mapping from the system state to the health indicator is not time-dependent. Based on this reasoning, time-independent and time-dependent deep learning models are considered in this research. In particular, we used two state-of-the-art deep neural networks as prognostics models: a deep multilayer perceptron feed-forward neural network (FNN) and a convolutional neural network (CNN). Deep CNN architectures have been proven to provide excellent performance on time series for predictive maintenance in recent works \cite{Li, Kiranyaz2019, Ellefsen2019}. Concretely, a solution based on 2D convolutions with 1D filters and no pooling layers was proposed in \cite{Li} that outperforms recurrent neural network (RNN) in the prognostics benchmark problem \cite{Saxena2008}. Similarly, deep architectures based on 1-d convolutions (see Figure \ref{fig:1-d_conv}) have also achieved excellent results in signal processing applications \cite{Kiranyaz2019}. Therefore, in light of the good results achieved, we adopted a One-dimensional convolutional neural network architecture as a time-dependent deep learning model.

To make the comparison for the proposed hybrid framework more challenging, we deliberately provided the purely data-driven approach an unfair advantage by selecting hyperparameters that maximize their prognostics performance while we do not perform hyperparameter tuning for the hybrid approach. We then train the same neural network models with the enhanced inputs space of the hybrid approach. Since the inputs space of hybrid and the purely data-driven approaches have a different dimension, the different sizes of the input changes also the number of weights in the first layer, and as a result, such architecture is sub-optimal for the hybrid approach. The network architectures of the deep learning prognostics model are briefly described below.

\textbf{Multilayer perceptron feed-forward neural network architecture (FNN)}. The architecture of the feed-forward neural network used in this research comprises five fully connected layers ($L=5$). The input layer has $n$ nodes, where $n$ denotes the size of the input space $\textbf{x}$ and varies depending on the considered solution strategy. The first three hidden layers have 200 neurons each. The last hidden layer has 50 neurons. A single linear neuron was used in the output layer. In compact notation, we refer to this architecture as $[n,\; 200,\; 200,\; 200,\;  50,\; 1]$. \textit{ReLU} activation function was used throughout the entire network. It should be noted that \text{RUL} estimation is a regression problem. Therefore, the last activation $\sigma^L=I$ is the identity. The network has 95k trainable parameters ($\mathcal{H}$). This final architecture is the result of conducting a grid-search where the search space over the  hyperparameters includes: number of hidden layers [1-4], number of neurons at each hidden layer [50, 100, 200] and activation function type [\textit{tanh}, \textit{relu}].

\textbf{One dimensional convolutional neural network architecture (1-d CNN)}. The architecture of the CNN neural network used in this research also comprises five layers ($L=5$). The network has three initial convolutional layers with filters of size $10$. The first two convolutions have ten channels and the last convolution has only one channel. Zero padding is used to keep the feature map through the network. The resulting 2D feature map is flattened and the network ends with a 50-way fully connected layer followed by a linear output neuron. The network uses \textit{ReLu} as the activation function. The network has 24k trainable parameters ($\mathcal{H}$). Similarly as for the FNN, the final architecture is the result of conducting a grid-search over the following hyperparameters: number of hidden layers [1-4], number of channels [10, 20, 30] at each convolutional layer, filter size [10, 20],  number of neurons at the fully connected layer [50, 100], activation function type [\textit{tanh}, \textit{relu}], and window size of the siding window [20, 50, 200].

\subsection{Training Set-up}
The optimization of the network's weights is carried out with mini-batch stochastic gradient descent (SGD) and with the \textit{AMSgrad} algorithm \cite{Kingma2014Adam}. \textit{Xavier} initializer \cite{Glorot} is used for the weight initializations. The batch size is set to 1024 and the learning rate to 0.001. The maximum number of epochs ($E_s$) was set to 60 for the FNN model and 30 for the CNN model. Early stopping with a patience of 5 epochs is considered.

\subsection{Evaluation Metrics} \label{sec:metrics}
The performance of the proposed framework is evaluated and compared to the purely data-driven deep learning models on the selected prognostics task. We apply two common evaluation metrics in C-MAPSS prognostics analysis \cite{Saxena2010}: root-mean-square error (RMSE) and NASA's scoring function ($s$) \cite{NASAdata} which are defined as:
\begin{align} \label{eq:score}
      s &= \sum_{j=1}^{m_*} exp(\alpha|\Delta^{(j)}|) \\
      RMSE &= \sqrt[]{\frac{1}{m_*}\sum_{j=1}^{m_*} {(\Delta^{(j)})}^2}
\end{align}
where $m_*$ denotes the total number of test data samples, $\Delta^{(j)}$ is the difference between the estimated and the real RUL of the $j$ sample (i.e., $y^{(j)}-\hat{y}^{(j)}$) and $\alpha$ is $\frac{1}{13}$ if RUL is under-estimated and $\frac{1}{10}$ otherwise. The resulting $s$ metric is not symmetric and penalizes over-estimation more than under-estimation.

Unit specific point-wise RUL estimation (i.e. $\hat{y}_{u}^{(j)}$) can show a high variability within a flight cycle, indicating that some parts of the flight are more informative for RUL estimation compared to other parts. In order to evaluate this effect we also define average RUL estimation at cycle $c$ in unit $u$ ($\hat{y}_u^{[c]}$) which is defined as follows: 
\begin{align}
	\hat{y}_{u}^{[c]} &= \frac{1}{m_u^{[c]}} \sum_{j=1}^{m_u^{[c]}} y_u^{(j)}
\end{align}
where $m_u^{[c]}$ is the length of the flight cycle $c$ for the \textit{u-th} unit; which is formally defined using the indicator function i.e. $\mathbf{1}\{.\}$ as:
\begin{align}
m_u^{[c]}=\sum_{j=1}^{m_*} \mathbf{1}\{U^{(j)} = u \; \wedge \; C^{(j)}=c\}
\end{align}
where $U$ and $C$ are vectors with unit and cycle labels for each sample of the test dataset.

\section{Experimental Results} \label{sec:Results}



\subsection{RUL Estimation} \label{sec:rul_estimation}
The RUL estimation is performed based on the same neural network models (i.e., FNN and CNN) and the same architectures for both setups: the purely data-driven and the proposed hybrid approach. In this section, the performance of the two approaches is compared based on different metrics. Table \ref{tb:compB} shows the performance on failure time prediction of the proposed hybrid approach ($\textbf{x} = [w, x_s, \hat{x}_s, \hat{x}_v, \hat{\theta}]$) and the \emph{baseline} approach (purely data-driven with $\textbf{x} = [w, x_s]$). With a reduction ranging from 16\% to $47\%$ in RMSE and $21\%$ to $68\%$ in $s$ score depending on the neural network model, the hybrid approach clearly outperforms the \emph{baseline}. Since the $s$-score penalizes over-estimation more than under-estimation and RMSE is a symmetric metric, the greater reduction of $s$ compared to RMSE indicates that the proposed method handles RUL over-estimation more effectively. The two neural network architectures evaluated reach nearly the same performance with the hybrid approach. However, important differences between time dependent and independent models based on a purely data-driven approach are observed (i.e., -38\% in RMSE between FNN and CNN). Overall, the hybrid approach with the CNN model is the best performing model.
\begin{table}[ht]
\begin{center}
\begin{tabular}{|l|c|c|c| }\hline
      \multicolumn{4}{|c|}{FNN model}                                                 \\\hline \hline
Metric              &  \text{Data-Driven} &  Hybrid                     &  rel. Delta \\\hline 
RMSE [-]            &  7.89 $\pm$ 0.12    &  $\mathbf{4.22}$ $\pm$ 0.10 &    -47\%    \\ \hline
$s \times 10^5$ [-] &  1.39 $\pm$ 0.04    &  $\mathbf{0.44}$ $\pm$ 0.01 &    -68\%    \\ \hline \hline
      \multicolumn{4}{|c|}{CNN model}                                                 \\ \hline \hline
Metric              &  \text{Data-Driven} &  Hybrid                     &  rel. Delta \\\hline 
RMSE [-]            &  4.95 $\pm$ 0.15    &  $\mathbf{4.14}$ $\pm$ 0.09 &    -16\%    \\ \hline
$s \times 10^5$ [-] &  0.56 $\pm$ 0.03    &  $\mathbf{0.44}$ $\pm$ 0.02 &    -21\%    \\ \hline
\end{tabular}
\end{center}
\caption[Table tb:compB]{Overview of the RMSE and $s$-score metrics with hybrid and baseline approaches for complete degradation trajectories on testset ($\mathcal{D}_{T*}$), comprising data from the three test units. Mean and standard deviation results with FNN and CNN models over five runs.}
\label{tb:compB}
\end{table}

\begin{figure}[ht]
\centering
\includegraphics[width=7.0cm]{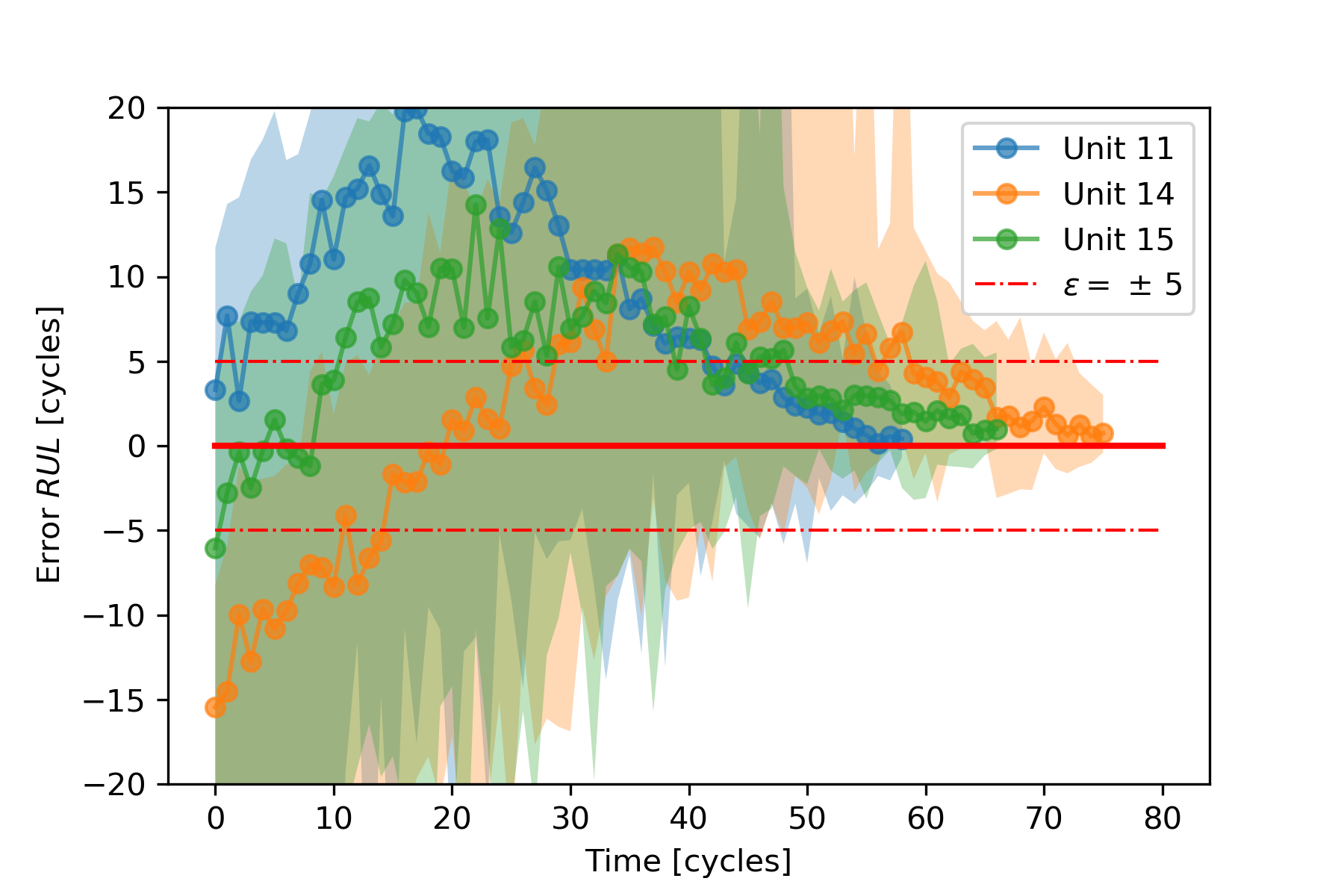}
\includegraphics[width=7.0cm]{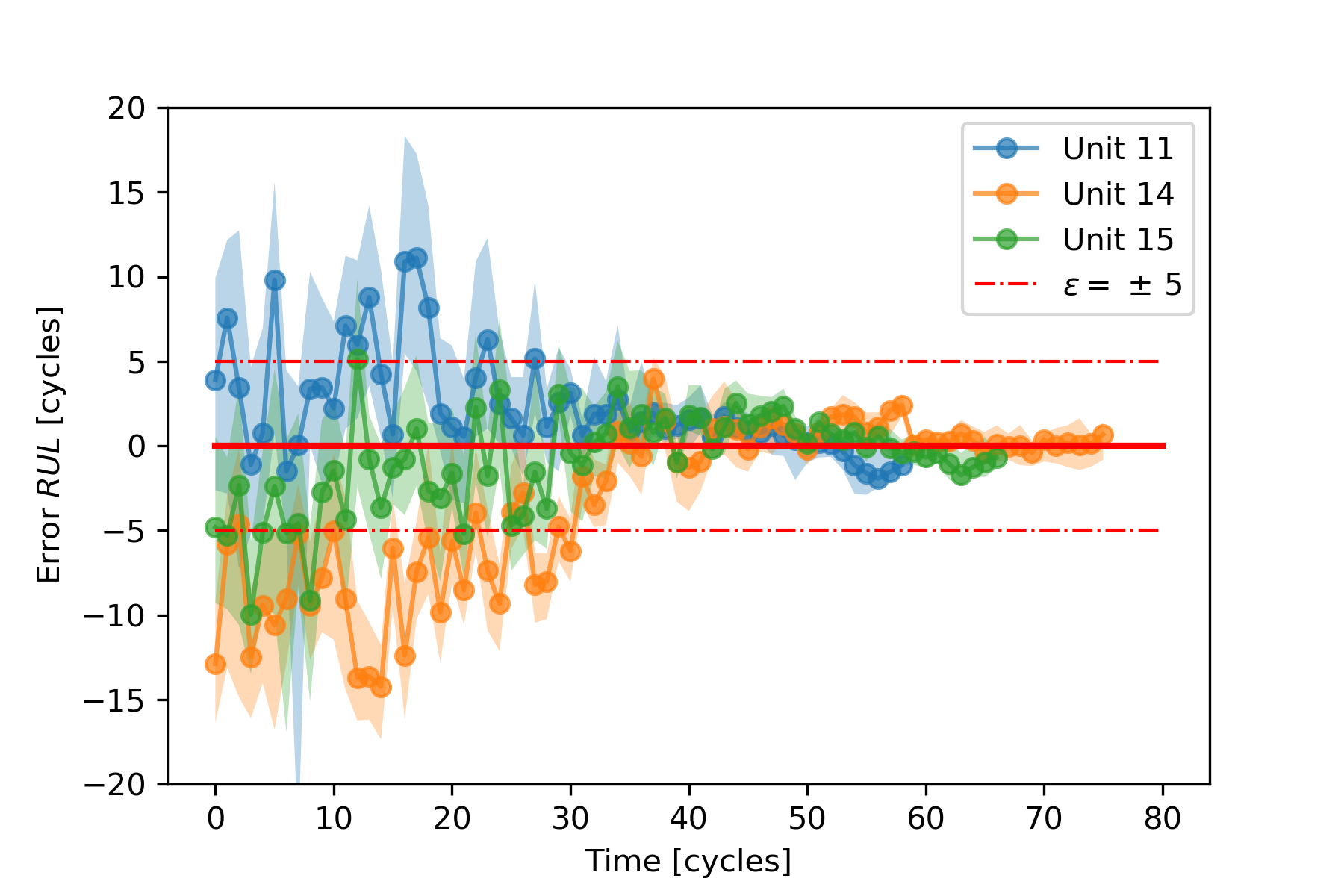}
\includegraphics[width=7.0cm]{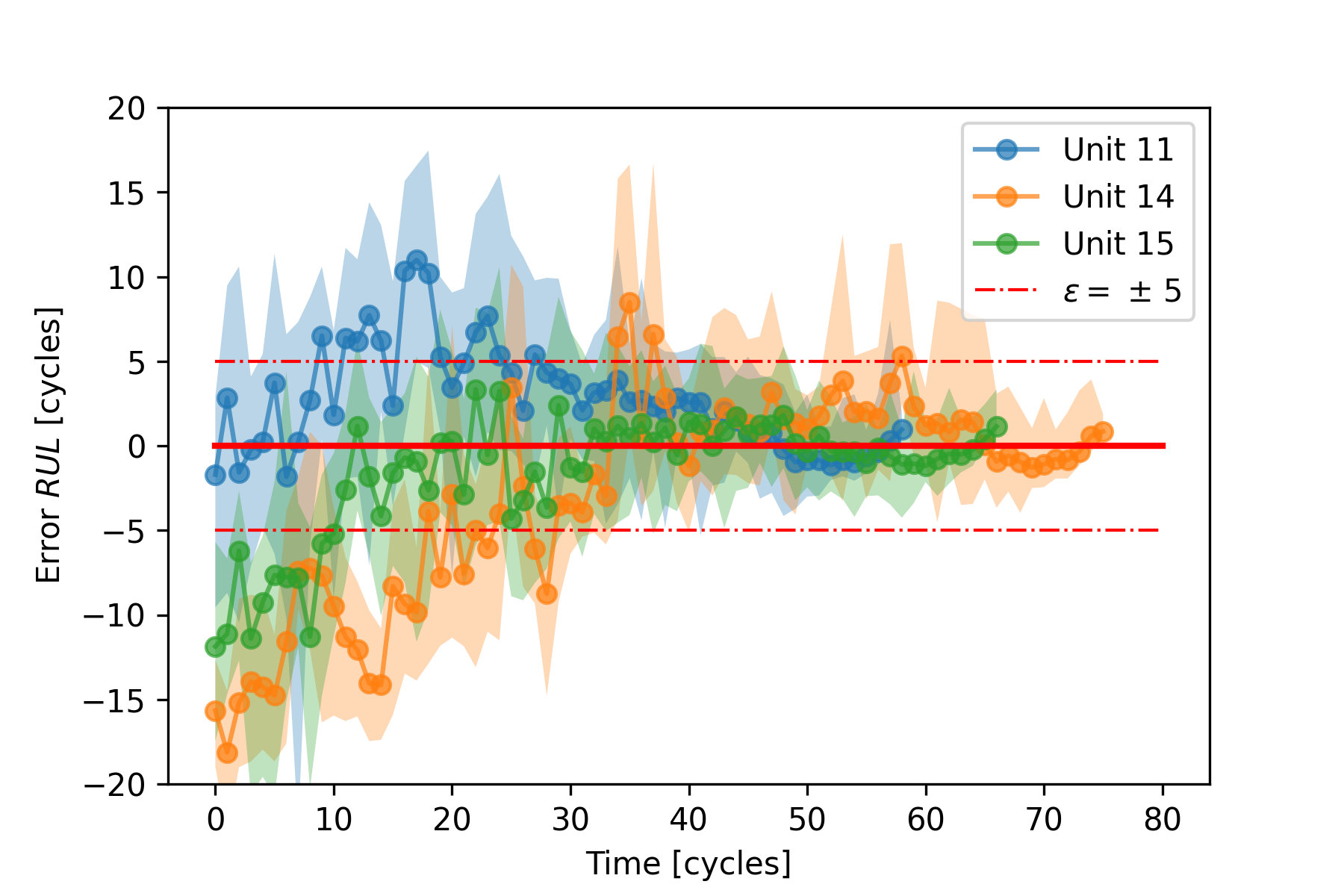}
\includegraphics[width=7.0cm]{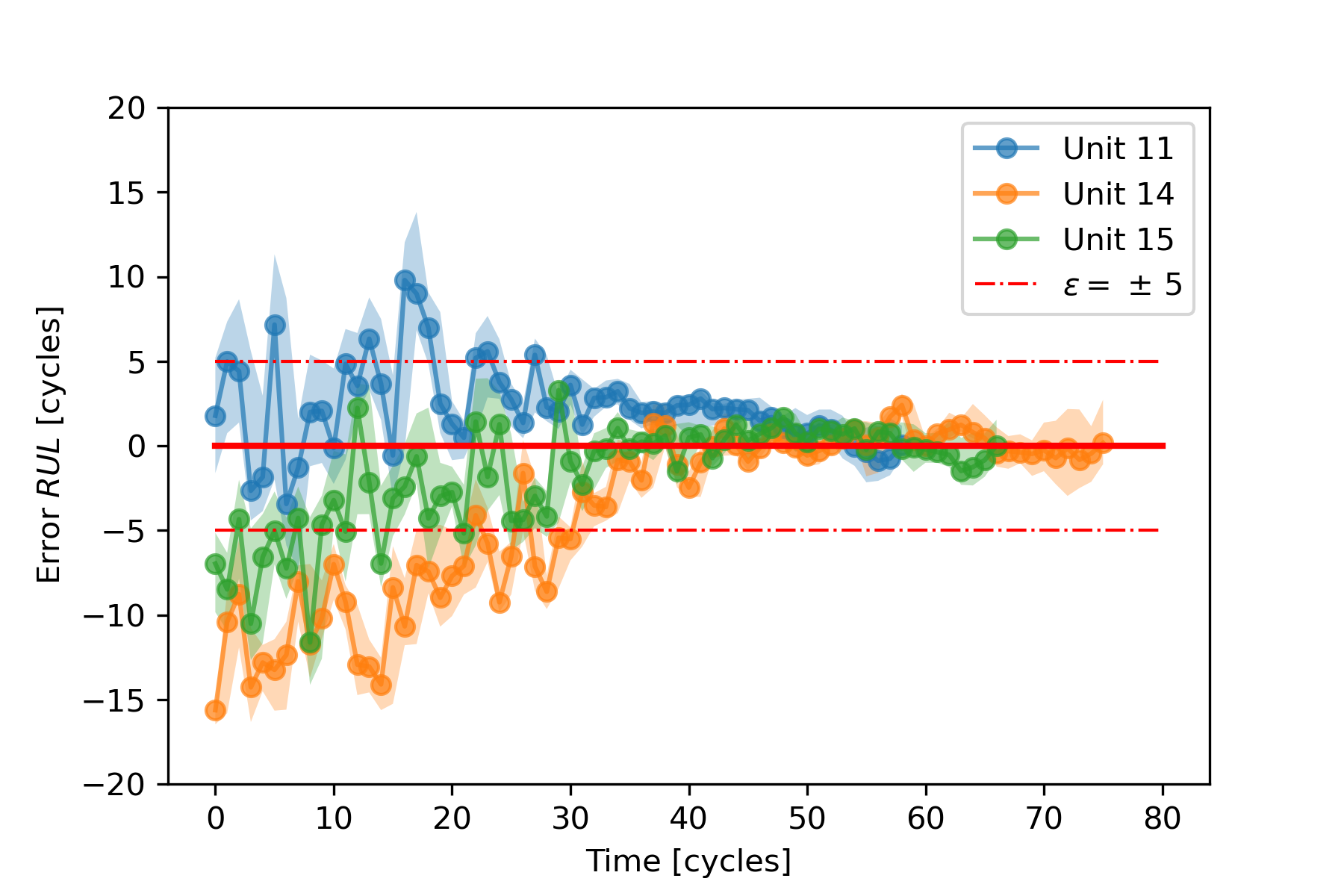}
\caption{RUL prediction error for each unit. The dotted lines are the average RUL estimate at each cycle i.e., $\Delta_{y_u} = \hat{y}_{u}^{[c]}-y_{u}^{[c]}$), the shaded surface is the uncertainty bounds for RUL predictions within each cycle. The red dashed horizontal lines correspond to $\pm 5$ cycles error bars. The three test units are shown: Unit 11 (blue), Units 14 (orange) and Unit 15 (green).}
 \label{fig:error}
\end{figure}

The improvement in prognostics performance is also observed for individual test units. Figure \ref{fig:error} shows the error between true and predicted \textit{RUL} (i.e., $\Delta_{y_u} = \hat{y}_{u}^{[c]}-y_{u}^{[c]}$) of the baseline approach (left) and the proposed hybrid approach (right) for each of the test units with FNN (top) and CNN models (bottom). The shaded surface shows the variability of the RUL predictions within each cycle. The upper limit corresponds to $\max(\Delta_{y_u})$ and the lower bound to $\min(\Delta_{y_u})$. \textit{RUL} estimates with the \emph{baseline} approach at any point in time have a large variability and high bias (specifically, RUL over estimation) compared to the hybrid approach. We can also observe that the cycle time where the prediction error is below 5 cycles for any future time ($t \geq c$) decreases for all the test units. We denote this time as $t_{|\Delta_y| \leq 5}$. Table \ref{tb:time_error_5} reports the prediction horizon i.e., $t_{\text{EOL}} - t_{|\Delta_y| \leq 5}$ for each unit and the average value for the whole fleet. Under this metric the proposed hybrid approach provides an average 127\% increase on the prediction horizon while maintaining a similar prediction accuracy.

\begin{table}[hbt!]
  \begin{center}
    \begin{tabular}{|c||c|c|c|}                                                     \hline
       $u$                     &  Data-Driven  & Hybrid           & rel.Delta    \\ \hline   
       11                      &  11           & $\mathbf{31}$    & 195\%        \\ \hline
       14                      &  15           & $\mathbf{43}$    & 197\%        \\ \hline
       15                      &  24           & $\mathbf{37}$    & 54\%         \\ \hline \hline
       Fleet Avg.              &  16           & $\mathbf{37}$    & 127\%        \\ \hline
    \end{tabular}
  \end{center}
   \caption[Table caption text]{Prediction horizon [cycles] for $\Delta_y \leq 5$ (i.e., $t_{\text{EOL}} - t_{|\Delta_y| \leq 5}$) with data-driven and hybrid approaches with CNN model.}
   \label{tb:time_error_5}
  \vspace{-0mm}
\end{table}

\subsection{Ablation Studies}
This section extends the previous analysis with three ablation studies to cover additional realistic prognostics scenarios and provide further insights into the proposed hybrid approach. Ablation study I evaluates the hybrid and the purely data-driven approaches under a variant of the case study with a smaller dataset. Ablation study II evaluates the impact of the calibration process quality on the hybrid approach's predictive performance. Finally, in ablation study III, we analyzed the contribution of the physics-derived features to the overall prediction performance.

\subsubsection{Ablation Study I: Impact of Dataset Size}
The results in the previous section showed that the proposed hybrid approach outperforms the purely data-driven approach independently of the neural network model used for the prognostics model. Moreover, it is worth noting that the CNN model provided a clear (i.e., 38\%) improvement in the performance over the FNN model. These results indicate that under abundant representative data, a powerful neural network model can serve as a competitive prognostics model. However, abundant run-to-failure datasets are often not available in real applications due to the rarity of occurring faults in safety-critical systems. Under scenarios of non-abundant CM data, we hypothesize that discovering informative and representative features of degradation from raw CM data is more challenging. As a result, the purely data-driven approach will have more difficulties in achieving good performance. To test this hypothesis, we consider an alternative dataset with a subset containing 50\% of the units. To decouple this analysis from the effect of dataset representatives (i.e., similar or dissimilar degradation trajectories), we selected units 16, 18, and 20 as they are affected by the same failure mode (i.e., HPT and LPT failure). The resulting prognostics performance of the purely data-driven and hybrid CNN models trained with training data from 50\% of the units are shown in Table \ref{tb:train_3_units}. The dataset size reduction has a negligible impact on the prognostics performance (2\% increase in RMSE) on the proposed hybrid approach. Contrarily, such a reduction of the dataset size leads to an increase of 17\% in the RMSE of the purely data-driven approach. The relative delta between hybrid and purely data-driven increases to 29\% in RMSE. Therefore, enhancing the input space with additional features from physical performance models becomes a clear advantage in scenarios of non-abundant CM data.  

\begin{table}[ht]
\begin{center}
\begin{tabular}{|l|c|c|c| }\hline
      \multicolumn{4}{|c|}{Units 2, 5, 10, 16, 18, 20}                                 \\ \hline \hline
Metric               &  \text{Data-Driven} &  Hybrid                     &  rel. Delta \\\hline 
RMSE [-]             &  4.95 $\pm$ 0.15    &  $\mathbf{4.14}$ $\pm$ 0.09 &    -16\%    \\ \hline
$s \times 10^5$ [-]  &  0.56 $\pm$ 0.03    &  $\mathbf{0.44}$ $\pm$ 0.02 &    -21\%    \\ \hline\hline
      \multicolumn{4}{|c|}{Units 16, 18, 20}                                           \\\hline \hline
Metric               &  \text{Data-Driven} &  Hybrid                     &  rel. Delta \\\hline 
RMSE [-]             &  5.97 $\pm$ 0.37    &  $\mathbf{4.22}$ $\pm$ 0.12 &    -29\%    \\ \hline
$s \times 10^5$ [-]  &  0.61 $\pm$ 0.05    &  $\mathbf{0.43}$ $\pm$ 0.02 &    -29\%    \\ \hline \hline
rel. Delta RMSE [\%] &  17\%               &    2\%                      &             \\ \hline
rel. Delta $s$ [\%]  &  8 \%               &    -2\%                     &             \\ \hline
\end{tabular}
\end{center}
\caption[Table tb:train3units]{Overview of the RMSE and $s$-score metrics with hybrid and baseline approaches for complete degradation trajectories on the testset ($\mathcal{D}_{T*}$), comprising data from the three test units. Mean and standard deviation results with CNN model over five runs. The training dataset contains only Units 16, 18 and 20}
\label{tb:train_3_units}
\end{table}

%

\subsubsection{Ablation Study III: Impact of Physics-Derived Information}
The proposed hybrid framework uses tree types of features derived from the calibrated physical-model, i.e., $\hat{x}_s, \hat{x}_v, \theta$. To analyzed the contributions of each of these feature types on the prognostics performance, we evaluated alternative hybrid models that increase the amount of model-derived information used in the prognostics model. In order to have a full coverage we evaluate the following combinations of input signals: the purely data-driven $[w, x_s]$, with added de-noised sensor readings $[w, x_s, \hat{x}_s]$, with added virtual sensors $[w, x_s, \hat{x}_v]$, and finally with added calibration factors $[w, x_s, \hat{x}_s]$. 

In order to decouple this analysis from the effect of uncertainly in the calibration process (see next section), we assume that the model information is obtained from a perfect calibration, i.e., $\hat{x}_s, \hat{x}_v, \theta$ are therefore ground-truth values. The resulting prognostics performances of the five new hybrid models are shown in Table \ref{tb:physics_impact}. We can observe that adding physics derived features always provides an improvement in the prediction performance. However, optimal performance is only achieved when the calibration factors are included in the prognostics model. This result suggests a hierarchy of information where the most informative features for RUL prediction, namely, the calibration parameters (i.e., $\theta$), are the most informative of the RUL. To verify this hypothesis, Figure \ref{fig:mi} shows the normalized mutual information between input features and the RUL target. The top-9 most informative input features within the input space $[w, x_s, \hat{x}_s, \hat{x}_v, \theta]$ are shown. The calibration parameters ($\theta$) are the most informative features for predicting RUL, followed by features representing operating conditions (i.e., $w$). The denoised sensor readings P50 and P2 (i.e., the total pressure at LPT outlet and the total pressure at fan inlet, respectively) are the most informative sensor.

\begin{minipage}{1.0\textwidth}
\begin{minipage}[b]{0.49\textwidth}
\centering
\includegraphics[width=6.5cm]{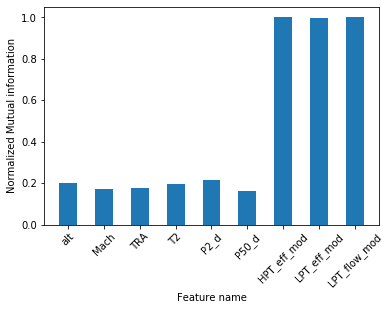}
\captionof{figure}{Top-9 features with higher normalized mutual information to the RUL label}
\label{fig:mi}
\end{minipage}
\begin{minipage}[b]{0.49\textwidth}
\centering
    \begin{tabular}{|l|c|c|}\hline  
    \multicolumn{3}{|c|}{ Purely Data-driven }                                         \\ \hline \hline
    Model                                     & RMSE [-]         &  $ s \times 10^5$   \\ \hline
    $[w, x_s]$                                &  4.95 $\pm$ 0.15 &   0.56 $\pm$ 0.03   \\ \hline
    \multicolumn{3}{|c|}{ Hybrid }                                                     \\ \hline \hline
    Model                                     & RMSE [-]         &  $ s \times 10^5$   \\ \hline
    $[w, x_s, \hat{x}_s]$                     &  4.90 $\pm$ 0.14 &   0.56 $\pm$ 0.03   \\ \hline
    $[w, x_s, \hat{x}_s, \hat{x}_v]$          &  4.61 $\pm$ 0.18 &   0.49 $\pm$ 0.05   \\ \hline
    $[w, x_s, \hat{x}_s, \hat{x}_v, \theta]$  &  4.14 $\pm$ 0.08 &   0.44 $\pm$ 0.02   \\ \hline
    \end{tabular}
  \captionof{table}{Overview of the RMSE and $s$ score results with CNN models under different levels of physics information. Mean $\pm$ standard deviation over 5 runs}
  \label{tb:physics_impact}
\end{minipage}
\end{minipage}

\subsubsection{Ablation Study IV: Impact of the Model Calibration Uncertainty}
\begin{figure}[ht]
\centering
\includegraphics[width=7.2cm]{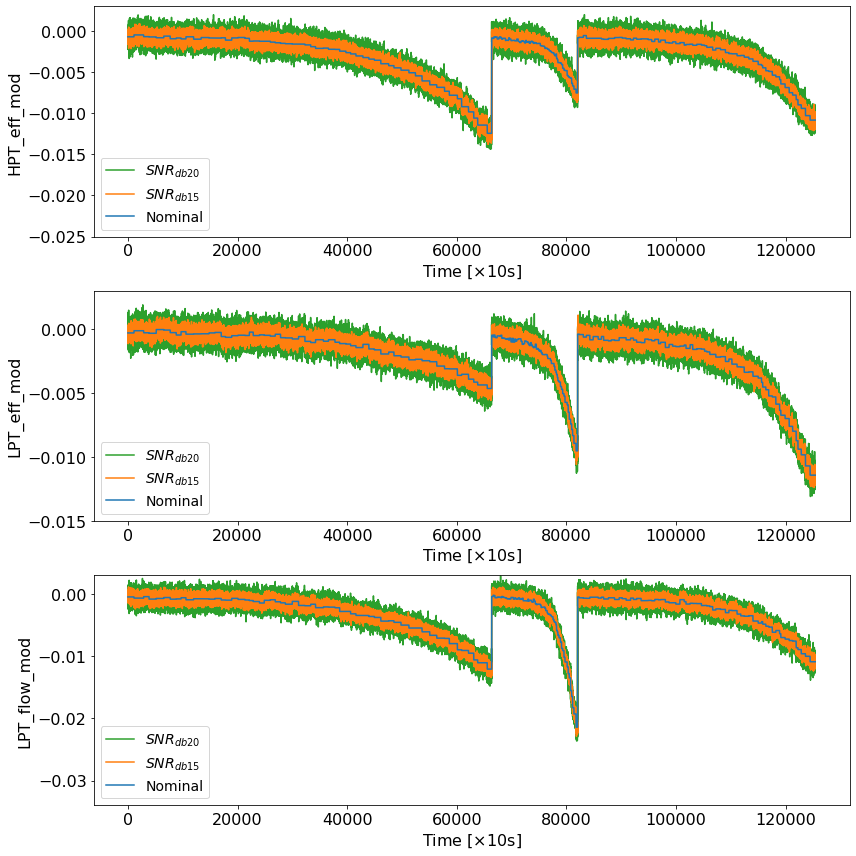}
\includegraphics[width=7.2cm]{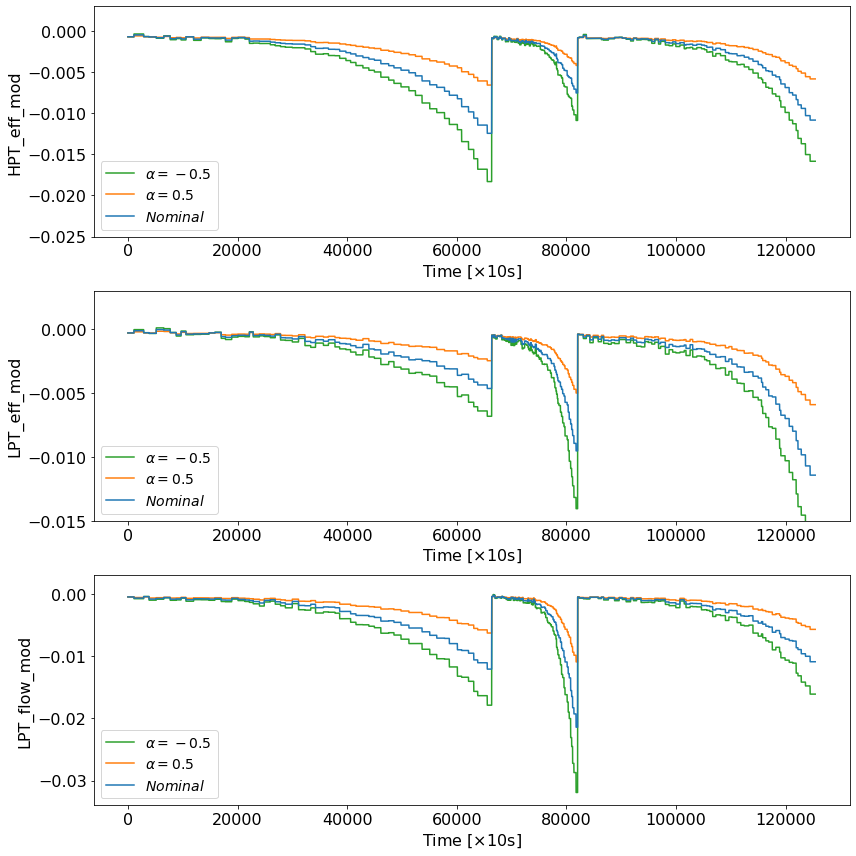}
\caption{\textbf{(Left)} Noisy calibration factors for a noise levels of $\text{SNR}_{db}=20$ (orange), and $\text{SNR}_{db}=15$ imposed on the high pressure turbine (HPT) efficiency, low pressure turbine (LPT) efficiency and flow.\textbf{(Right)} Biased calibration factors with increasing shift with $\alpha=-0.5$ and $\alpha=0.5$. Both plots are the $\theta$ values for the three units are stacked one after the other, generating a single time sequence. }
 \label{fig:noise_signal}
\end{figure}

The quality of the calibration process has an impact on the prognostic performance of the proposed hybrid framework. In order to quantify this impact, we evaluate the sensitivity of the proposed framework to the calibration performance, i.e. the impact of low quality estimates of $\theta$ on the RUL prediction performance. Concretely, we consider hypothetical situations where $\hat{\theta}$ is noisy (see Figure \ref{fig:noise_signal}, left) and also where $\hat{\theta}$ is affected by bias (see Figure \ref{fig:noise_signal}, right). To model the noisy calibrations of different quality, white noise with signal-to-noise ratios $\text{SNR_{db}}$ of 15 and 20 db are imposed on the calibration factors (i.e., \emph{HPT\_Eff\_mod}, \emph{LPT\_Eff\_mod} and \emph{LPT\_Flow\_mod}). To model the bias, we imposed a increasing shift proportional to the nominal calibration factor values ($\theta_{\alpha} = \theta^{(t)}  + \alpha( \theta^{(0)} -\theta^{(t)})$. Two values of $\alpha$ are evaluated (i.e., $\alpha=0.5$ and $\alpha=-0.5$) representing a $\pm$50\% error in the estimation of  $\theta$. Since $\hat{\theta}$ is multidimensional and an infinite number of scenarios is possible, we restricted the analysis to case where the calibration factors are affected equally by noise and bias. Figure \ref{fig:noise_signal} (left) shows the resulting noisy calibration parameters $\hat{\theta}$. The right figure shows the biased calibration factors.
Table \ref{tb:robust_track} reports the impact in RMSE and $s$ of the three $\text{SNR}$ evaluated values. We can observe a decrease in the accuracy as the noise increases. However, even with a very high SNR of 15, the proposed hybrid framework is still able to achieve a better prognostics performance than the purely data-driven model. Therefore, these results demonstrate the robustness of the proposed hybrid prognostics framework. 

\begin{table}[ht]
  \begin{center}
    \begin{tabular}{|c||c|c|} \hline
      \multicolumn{3}{|c|}{ Bias }                                      \\ \hline \hline
       \text{Intensity}         & RMSE [-]          &  $ s \times 10^5$ \\ \hline
       $\alpha =  +0.5$          & 4.77 $\pm$ 0.24  &  0.47 $\pm$ 0.02  \\ \hline 
       $\alpha =  -0.5$          & 4.66 $\pm$  033  &  0.59 $\pm$ 0.07  \\ \hline \hline 
    \multicolumn{3}{|c|}{ Noise}                                        \\ \hline \hline
       \text{Intensity}         & RMSE [-]          &  $ s \times 10^5$ \\ \hline
       $\text{SNR}_{db}=20$     &  4.26 $\pm$ 0.13  &  0.47 $\pm$ 0.03  \\ \hline
       $\text{SNR}_{db}=15$     &  4.71 $\pm$ 0.17  &  0.52	$\pm$ 0.02  \\ \hline
    \end{tabular}
  \end{center}
\caption[Table tb:time]{Overview of the prognostics performance under model bias and noise with hybrid approaches and 1-dCNN model.}
\label{tb:robust_track}
\end{table}

\section{Conclusions} \label{sec:Conclusion}

The work presented in this paper proposes a hybrid framework fusing information from the physics-based performance models with deep learning algorithms for predicting the remaining useful lifetime of complex systems. Health-related model parameters are inferred by solving a calibration problem. Subsequently, this information is combined with sensor readings and used as input to a deep neural network to develop a reliable hybrid prognostics model. 

The performance of the hybrid framework was evaluated on a synthetic dataset comprising run-to-failure degradation trajectories from a small fleet of nine turbofan engines operating under real flight conditions. The dataset was generated with the Commercial Modular Aero-Propulsion System Simulation (C-MAPSS) dynamical model. The proposed hybrid framework clearly outperforms the alternative data-driven prognostic model where only sensor data are used as input to the deep neural network. The hybrid framework provides accurate and robust predictions of the failure time, $t_{\text{EOL}}$ (i.e., $\Delta \pm 5$ cycles) while extending the prediction horizon by 127\% on average compared to the pure data-driven methods. More importantly, the proposed framework requires less training data compared to the purely data-driven algorithms. 

As demonstrated in the experiments, the performance of the proposed hybrid prognostics framework is robust to uncertainty imposed by the quality of the model calibration. This research study has also demonstrated the capabilities of the developed hybrid framework to provide excellent prognostics performance for units that exhibited very dissimilar operating conditions compared to the operating conditions of units used for training the algorithms (limited representativeness of the training dataset for the testing dataset). Therefore, the proposed hybrid framework sets a promising direction for further research in PHM applications.

Finally, a potential future research direction is evaluating the transferability of the proposed hybrid approach to other types of problems that fulfill the same criteria: availability of complete physics-based models and the availability of sensor readings that provide information about the system state. Furthermore, a further future research direction is developing approaches for which only limited physics-based information is available. 

\section*{Acknowledgements} \label{sec:acknowledgements}

This research was funded by the Swiss National Science Foundation (SNSF) Grant no. PP00P2 176878.

\bibliographystyle{unsrt}
\bibliography{reference}

\end{document}